\documentclass[sigplan,screen,nonacm]{acmart}

\IfFileExists{libertine.sty}{}{\usepackage{lmodern}}

\usepackage{xcolor}
\usepackage{graphicx}
\usepackage{tabularx}
\makeatletter
\let\resolve@savedpath\path
\makeatother
\usepackage{tikz}
\usetikzlibrary{arrows.meta,positioning,shapes.geometric,fit,backgrounds,matrix,calc}
\makeatletter
\let\path\resolve@savedpath
\makeatother
\definecolor{darkgreen}{HTML}{006400}
\newcommand{\red}[1]{\textcolor{red}{#1}}
\newcommand{\green}[1]{\textcolor{darkgreen}{#1}}

\newcommand{\nfull}{14}

\ifdefined\INTROONLY
\makeatletter
\def\resolve@setref#1#2#3{%
	\ifx#1\relax\textcolor{red}{[later]}\else\expandafter#2#1\fi}
\AtBeginDocument{\let\@setref\resolve@setref}
\makeatother
\fi

\setcopyright{none}
\begin{document}

\title{RESOLVE: Language-Agnostic Validation of GPU Kernels
	Through Testing, Reduction, and Proof}

\author{Ashkan Vedadi Gargary}
\affiliation{%
	\institution{University of California, Riverside}
	\city{Riverside}
	\state{California}
	\country{USA}
}

\author{Guido Mart{\'i}nez}
\affiliation{%
	\institution{Math, Inc.}
	\city{New York City}
	\state{New York}
	\country{USA}
}

\author{Sebastian Burckhardt}
\affiliation{%
	\institution{Microsoft Research}
	\city{Redmond}
	\state{Washington}
	\country{USA}
}

\author{Gabriel Ebner}
\affiliation{%
	\institution{Microsoft Research}
	\city{Redmond}
	\state{Washington}
	\country{USA}
}

\author{Abhinav Jangda}
\affiliation{%
	\institution{Microsoft Research}
	\city{Redmond}
	\state{Washington}
	\country{USA}
}

\author{Madan Musuvathi}
\affiliation{%
	\institution{Microsoft Research}
	\city{Redmond}
	\state{Washington}
	\country{USA}
}

\author{Tyler Sorensen}
\affiliation{%
	\institution{Microsoft Research}
	\city{Redmond}
	\state{Washington}
	\country{USA}
}

\renewcommand{\shortauthors}{Vedadi Gargary et al.}

\begin{abstract}
AI systems can now write and optimize production GPU kernels, 
but validating them remains an important challenge. Evaluating the kernel 
on a few random inputs and checking that its outputs
match a trusted reference kernel within numeric tolerances is not sufficient:
races can cause nondeterministic behavior that fails to manifest in tests, and 
numeric tolerances can hide bugs and cause false positives
even after extensive calibration.

To address this challenge, we present RESOLVE,
which combines testing and formal verification to build a comprehensive
kernel validation pipeline. It operates in three steps:
First, it tests for nondeterminism using binary instrumentation that perturbs execution
timing to expose races. 
Second, an agent rewrites the candidate and reference kernels 
to obtain "reduced-concurrency" versions 
that are simpler to analyze but still produce 
bitwise-identical outputs in all tests.
Third, the reduced kernels are formally analyzed in the F*/Pulse framework
and prove that they perform 
the same computation on real numbers. This sidesteps the need for numeric tolerances.
We show that RESOLVE can validate a broad
selection of kernels using KernelBench, and prove equivalence across fused GEMMs in
three state-of-the-art frameworks and languages: CUTLASS, Triton, and Gluon.
It also analyzes mega-kernels, notoriously difficult to validate, and finds four
previously unreported issues, including two clear bugs.  We show that agents can use RESOLVE to repair the issues, with minimal performance impact, highlighting that agents can optimize aggressively
when they can rigorously check their results.
\end{abstract}

\maketitle

\section{Introduction}

GPU kernels determine much of the performance of modern AI workloads. Recent systems have shown that AI agents can
generate and optimize these kernels~\cite{Ouyang2025KernelBench,Li2025TritonBench,Ye2026CAKE}, producing candidates far
faster than humans can review them for correctness. Kernel generation is
therefore creating a new validation bottleneck: an optimization that
cannot be trusted cannot safely be deployed, regardless of its speed.

The predominant method for validating GPU kernels is reference-based
testing~\cite{Ouyang2025KernelBench,Li2025TritonBench}: execute a
candidate and a trusted reference on representative inputs and compare
their outputs. This approach is often effective for machine-learning
kernels, whose behavior is determined largely by tensor shapes, data
types, layouts, and other discrete configurations that can be sampled
systematically. Nevertheless, two challenges remain even with broad input coverage.

First, concurrency bugs may manifest only under rare thread or warp
schedules and therefore remain hidden across repeated executions.
Second, valid optimizations may reorder floating-point operations and
produce outputs that are not bitwise identical to the reference.
Testing must consequently compare outputs using a numerical tolerance,
but a permissive tolerance can conceal bugs while a restrictive one can
reject valid optimizations.

Formal verification can instead establish equivalence for all inputs
covered by a specification~\cite{Martinez2026Kuiper,
Chatterjee2026ProofWright}. Its applicability, however, is limited by
the verifier's semantic coverage. Modern kernels rely on warp
intrinsics, asynchronous memory operations, tensor-core instructions,
and synchronization patterns that existing verification frameworks do
not fully model. Because GPU programming models and hardware features
evolve faster than their formalizations, directly verifying optimized
production kernels remains impractical.

This paper presents RESOLVE, a GPU-kernel validation framework that
bridges the gap between the broad applicability of testing and the
rigorous guarantees of formal verification. RESOLVE's central idea is
to separate \emph{concurrency correctness} from
\emph{functional correctness}, and to use a different validation
technique for each. This decomposition exploits a key property of
machine-learning kernels: they are generally intended to be
deterministic, producing the same output for a given input regardless of
thread schedule. This allows RESOLVE to focus on capturing a broad set of concurrency issues as schedule-dependent nondeterminism, while deferring functional equivalence to a formal proof.

Given a candidate kernel $C$ and a trusted reference kernel $R$,
RESOLVE asks agents to generate reduced-concurrency kernels $C_r$ and
$R_r$. These reductions remove asynchronous pipelines, synchronization,
and other concurrency constructs that impede formal reasoning, while
seeking to preserve the bitwise behavior of their respective originals. Thus,
$C$ and $R$ may use complex concurrency patterns and recent hardware
features, while $C_r$ and $R_r$ are designed to fall within the
semantic coverage of the verifier.

RESOLVE validates concurrency correctness using systematic concurrency
testing~\cite{Burckhardt2010PCT,Stephenson2026SuperCollider}. Using the NVBit
binary-instrumentation framework~\cite{Villa2019NVBit}, RESOLVE perturbs
the timing of scheduling-relevant operations to explore different schedules. A lightweight memory analysis focuses these
perturbations on GPU scopes and memory regions in which conflicting
accesses may occur. As this analysis operates on the compiled NVIDIA GPU binary, it is
independent of the source language or framework used to generate the
kernel, including CUDA, Triton~\cite{Tillet2019Triton},
Gluon~\cite{Triton2026Gluon}, and CUTLASS~\cite{NVIDIA2023CUTLASS}.

RESOLVE then compares executions bit-for-bit. For a fixed input,
repeated executions of each kernel must produce identical results, and
each reduced kernel must produce the same result as its original. These
comparisons distinguish two failure modes. If repeated executions of
the same kernel differ, RESOLVE reports schedule-dependent behavior. If
an original and its reduced kernel are individually deterministic but
differ from one another, RESOLVE rejects the reduction and asks the
agent to try again. At these boundaries, bitwise comparison provides an
exact per-execution oracle and requires no numerical tolerance.

To verify functional correctness, agents translate $C_r$ and $R_r$
into Kuiper~\cite{Martinez2026Kuiper} kernels $C_k$ and $R_k$, respectively.
Kuiper is a verification language built on the F*/Pulse concurrent separation
logic~\cite{Ebner2025PulseCore} that covers a subset of CUDA and compiles to it.
RESOLVE checks the correspondence
between each Kuiper translation and its reduced executable using
bitwise testing. Agents then construct an algebraic specification from
the trusted reference alone and prove in Kuiper that both $C_k$ and
$R_k$ satisfy it. Because the specification is derived without
consulting the candidate, the candidate cannot influence the criterion
against which it is judged. A successful proof establishes that $C_k$
and $R_k$ implement the same algebraic computation, even when valid
floating-point reorderings cause them to produce numerically different
results.

RESOLVE therefore uses testing where bitwise equality provides an exact
per-execution oracle and formal verification where numerical testing
requires an approximate one. Testing connects optimized binaries to
proof-friendly representations, extending formal reasoning to kernels
that use languages and hardware features outside the verifier's
semantic coverage. Agents construct the reductions, translations,
specifications, and proofs, but do not serve as correctness oracles:
their outputs must pass bitwise testing or formal proof checking.

We evaluate RESOLVE on KernelBench kernels, complex state-of-the-art fused GEMM kernels written using Triton, Gluon, and CUTLASS, and three mega-kernels. Although every KernelBench candidate had passed the benchmark's existing tolerance-based tests, RESOLVE identified two incorrect candidates. We also design a mutation testing strategy to evaluate how effectively our concurrency testing finds synchronization faults. The fused GEMMs in particular use concurrency and hardware features outside Kuiper's directly supported subset, and RESOLVE validated them regardless, demonstrating that reduction can extend formal reasoning to kernels that cannot be verified directly. Across the mega-kernels, RESOLVE found four previously unreported issues, including two clear bugs. Agents repaired them, and RESOLVE tested the repairs, which changed performance by less than 2\% on three of the four.

\paragraph{Contributions.}
In summary, we present RESOLVE, a workflow to validate GPU kernels.  It
consists of:

\begin{enumerate}\setlength{\itemsep}{1pt}\setlength{\parskip}{0pt}
	\item \textbf{Efficient binary-level determinism testing}
	      (\S\ref{sec:phase-dtester}): timing perturbation on SASS,
	      needing no source or compiler and covering any language or library.
	\item \textbf{Bitwise concurrency reduction} (\S\ref{sec:reduction}): an agent
	      rewrites both programs with as little concurrency as possible,
	      accepted only when the output is bitwise equivalent.
	\item \textbf{Automated equivalence proof} (\S\ref{sec:kplay}): agents write
	      a specification from the trusted reference alone and a program for
	      each side satisfying it, so an optimization can differ bitwise and
	      still be accepted.
	\item \textbf{Evaluation} (\S\ref{sec:eval-breadth}): sampled KernelBench
        kernels, state-of-the-art fused GEMMs in Triton, Gluon, and CUTLASS,
        and three mega-kernels, in which we find four unreported issues.
\end{enumerate}

Given those reports, agents repaired the bugs, four with verified deterministic
fixes, showing that better testing tools can make agentic kernel optimization
more ambitious.

\ifdefined\INTROONLY
\bibliographystyle{ACM-Reference-Format}
\bibliography{reference}
\end{document}
\fi

\section{GPU Background and Setup}

A GPU is built from many simple cores and gets its performance from running
enormous numbers of threads.  Recent GPUs also have fixed-function \emph{tensor
cores} that perform a small matrix multiply in a single instruction.  A GPU
function is called a \emph{kernel}, and many thousands of \emph{threads} run it
at once.  Threads are grouped into \emph{warps} of 32 that advance together,
warps into \emph{cooperative thread arrays} (CTAs, also called blocks) whose
threads share a fast on-chip scratchpad, and CTAs into a \emph{grid} that covers
the device.  The newest hardware adds a \emph{cluster} level between the CTA and
the grid.  Memory follows the same levels: each thread has private
\emph{registers}, each CTA has \emph{shared memory} that only its own threads
can reach, and every thread can reach a large \emph{global memory}.  Launching a
kernel runs one grid to completion.

Most kernels are still written in CUDA, but a lot of performance work now
happens in languages above it.  Triton~\cite{Tillet2019Triton} lets a kernel be
written over tiles and leaves the compiler to assign work to threads and to
place data in shared memory.  Gluon~\cite{Triton2026Gluon} sits on the same
compiler but exposes layouts, memory movement and asynchrony, so the programmer
controls what Triton decides.  CUTLASS~\cite{NVIDIA2023CUTLASS} is a C++
template library of hand-tuned building blocks for matrix multiply and its
epilogues.
Vendor libraries such as cuBLAS and cuDNN ship as binaries, with no source or
PTX to inspect.  The only common foundation is \emph{SASS}, the relatively
undocumented binary code for NVIDIA GPUs, unique to each architecture.

Kernels have been getting larger.  A single operation is usually limited by
memory rather than arithmetic: it reads its inputs from global memory, computes,
and writes the result back for the next operation to read again.  Fusing several
operations into one kernel removes those round trips.  Fusion used to mean a
matrix multiply with an activation attached to it.  Now it means the whole
\emph{MLP} block of \S\ref{sec:eval-mlp}, the multi-layer perceptron that
follows attention in every transformer layer of a modern LLM, or the
mega-kernels of \S\ref{sec:eval-depth}, which put an entire model in one launch.
Two of those mega-kernels implement a \emph{mixture of experts}, which replaces
the MLP block with many copies of it and routes each token to a few.

Fusing more work into one kernel puts more concurrency in it, and spreads that
concurrency across more scopes.  This paper is concerned with \emph{races}:
memory conflicts, at least one of them a write, unordered by synchronization.  A
race gives a different outcome depending on the scheduler.  Races can occur at
every level of the hierarchy, and each level has its own way of ordering
accesses: threads in a warp advance together, threads in a CTA can be ordered
with \texttt{\_\_syncthreads()}, and separate CTAs need a grid-wide barrier or
atomics.  The memory scopes match, so shared memory can only be raced on from
inside one CTA and global memory from anywhere.  Asynchronous copies and TMA
transfers add another boundary, between issuing a transfer and the point at
which its result is safe to read.

Reasoning about GPU programs almost always requires reasoning about
floating-point computation, which does not behave like real arithmetic.
Addition is not
associative, so summing the same values in a different order changes the last
bits, and any kernel that splits a reduction across threads is choosing an
order.  Fast-math flags widen the gap by letting the compiler reassociate,
contract a multiply and an add into one fused operation, and replace a division
by a reciprocal.  Transcendental functions have no exact implementation, so an
exponential ends up as a polynomial or a Taylor series, and two kernels can
choose different ones.  Tensor cores accumulate in their own precision and their
own order, so a matrix multiply run on them does not agree bit for bit with the
same multiply built out of ordinary fused multiply-adds.  Two kernels that
compute the same mathematical function may not produce the same bits.

The kernels in this paper are tensor kernels, and we assume their shapes are
fixed.  ML kernels are tuned, and often recompiled, for the shapes they will
actually see, so this is how they are written anyway.  It also makes our testing
tractable: with the shapes fixed, a kernel issues a fixed number of memory
accesses, so it can be traced in full and replayed on the same input under many
schedules.

\subsection{RESOLVE Kernel Bench and Kernel Executors}
\label{sec:setup}

Throughout this work we need kernels to evaluate, and we draw on several
corpora.  For breadth and characterization we built one, the RESOLVE Kernel
Bench (RKB), which pairs each of the 100 KernelBench Level~1 kernels
\cite{Ouyang2025KernelBench} with an optimized candidate.  We generated the
candidates with GPT-5.6 Sol and retried until each compiled and passed the
tolerance tests shipped with the benchmark, so every candidate has already
passed a conventional tolerance oracle.

RESOLVE assumes each kernel comes with a kernel executor.  It runs the kernel a
fixed number of times and hashes all of memory after each run, not just the
output.  That hash is what every comparison in RESOLVE is made on, between two
runs of the same kernel and between a kernel and its reduced program.  Hashing
is only a way to save space: the comparisons are all equality tests, so we never
need to keep a full copy of memory from every run.

Inputs are tensors drawn from ten families of distributions, chosen per
operation and seeded so a run can be replayed on the same input and any
difference between runs is a property of the kernel.  Uniform random data can
miss branches and leave corner cases of floating-point arithmetic unexercised,
so the families are built to cover both: degenerate, sparse and sentinel values,
awkward magnitudes and cancellation, quantization boundaries, and the ties,
indices, masks and lengths that steer control flow.  We give their definitions
in the artifact.  Every kernel is run a fixed number of times on each input.

\section{Testing for Determinism}
\label{sec:determinism}
\label{sec:phase-dtester}

RESOLVE requires the kernels it validates to be deterministic: on a given
input, a kernel must produce the same output on every run. The \emph{determinism
tester} checks this, treating the kernel as a black box. Requiring determinism
has two big
advantages. First, it enables and improves the rest of our reduction pipeline,
by simplifying subsequent reductions and boosting the effectiveness of random
testing. Second, it can flag synchronization errors in kernels. For example, a
mistakenly omitted barrier can cause nondeterminism.

Requiring determinism also has a potential disadvantage: we may
reject kernels that are non-deterministic and yet correct (\emph{legitimate nondeterminism}).
For example, a kernel may intentionally use a nondeterministic reduction order
(say, based on \texttt{atomicAdd}), leading to
slightly different results due to the non-associativity of floating-point addition.

Empirically, legitimate nondeterminism is rare for GPU kernels. Only one of the
100 kernels in RKB uses a floating-point \texttt{atomicAdd}; the rest appeared
deterministic on inspection and under testing (\S\ref{sec:eval-breadth}), so the
benefits outweigh the disadvantage.

\begin{figure}[t]
	\centering
	\begin{tikzpicture}[
			font=\footnotesize,
			pass/.style={draw, align=center, text width=58pt,
					inner xsep=4pt, inner ysep=4pt},
			flow/.style={-{Stealth[length=4pt]}, draw=black!70, line width=0.5pt},
			ask/.style={align=center, text width=66pt, font=\footnotesize\itshape},
		]
		\node[pass] (c) {counting};
		\node[pass, right=10pt of c] (t) {tracing};
		\node[pass, right=10pt of t] (p) {perturbation};
		\draw[flow] (c) -- (t);
		\draw[flow] (t) -- (p);
		\node[ask, below=5pt of c] {how much memory traffic is there?};
		\node[ask, below=5pt of t] {where can a conflict be?};
		\node[ask, below=5pt of p] {does the schedule change the output?};
	\end{tikzpicture}
	\caption{The three passes RESOLVE makes over the binary, and the question
		each one answers.}
	\label{fig:passes}
\end{figure}
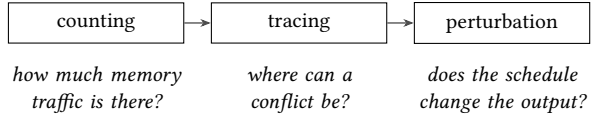

\paragraph{Determinism vs. Race-Freedom} The two often occur in tandem, but they
are defined differently and neither implies the other. Data races are
\emph{violations of synchronization rules} agreed upon by a community, and there
are many definitions rather than one universally accepted one. Evidence for a
data race must include execution-internal details to show the violation.
Nondeterminism is externally observable, and two executions that produce a
different output on the same input suffice as evidence.

\subsection{Characterization}

RESOLVE looks for races by delaying memory operations and comparing the
kernel's full output bitwise against an unperturbed run.  Delaying an access
that races with another changes which of the two lands first, and the kernel
can produce a different result.  SuperCollider
\cite{Stephenson2026SuperCollider} applies the same idea to individual memory
accesses, which it reaches through a compiler pass.

The difficulty is knowing
where to delay: a GPU kernel runs enormous numbers of threads over large
tensors, and delaying at random took hours per kernel in our pilot
runs.  Conflicting memory accesses are not spread evenly, though: whole memory
regions and levels
of the thread hierarchy may admit none at all. Since determinism bugs
do not occur without conflicts, there is no point in testing such accesses.
RESOLVE therefore measures the kernel in three
passes over the binary, counting, tracing, and perturbation
(Figure~\ref{fig:passes}), placing delays only where the trace shows a conflict
is possible.  This brings the median time to test a kernel in RKB down to 88
seconds (\S\ref{sec:eval-breadth}).

\paragraph{How much memory traffic is there?}
\label{sec:counting}
Tracing memory accesses requires a buffer, and that buffer has to be sized
before the kernel runs, either to fit in memory or to keep the later analysis
tractable.  The first pass therefore runs the kernel under NVBit
(NVIDIA's binary instrumentation tool) and counts how
many memory accesses each CTA (cooperative thread array)
 performs, along with the grid shape and the
number of TMA (tensor memory accelerator) descriptors.  Because nothing is written to the buffer and no
analysis runs, the cost is instrumentation and counting alone.  CTAs are
sampled at random, so that later analysis is not biased toward the blocks with
the lowest identifiers.

\begin{figure}[t]
	\centering
	\includegraphics[width=.98\columnwidth]{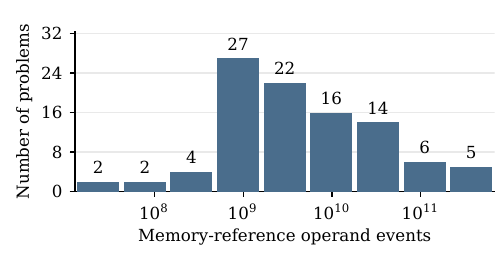}
	\caption{Memory-reference operand events per problem across RKB, measured by
		the counting pass on H100.  Activity spans roughly four orders of
		magnitude, from under $10^{8}$ to beyond $10^{11}$.}
	\Description{Histogram over the 98 RKB problems with a nonzero count, on a
		logarithmic axis of memory-reference operand events per problem.  Bar
		heights from left to right are 2, 2, 4, 27, 22, 16, 14, 6, and 5 problems,
		peaking just below ten to the ninth and tailing off past ten to the
		eleventh.}
	\label{fig:traffic}
\end{figure}

Figure~\ref{fig:traffic} shows the distribution of memory-reference operand
events per problem across RKB.  Activity spans roughly four orders of
magnitude: most problems fall between $10^{8}$ and $10^{10}$, while the largest
reach beyond $10^{11}$.  Two problems, P31 and P32, count zero memory-reference
operands and are excluded from the figure; \S\ref{sec:eval-breadth} returns to
them.  These counts let RESOLVE scope the
trace buffer and the analysis to each kernel, instead of over-provisioning for
one that needs little or failing on one that needs more than expected.  They
also make explicit what any later sampling leaves out.

\paragraph{Where can a conflict be?}
\label{sec:tracing}
The second pass records the memory accesses themselves, including each address
and the thread that issued it.  From this trace RESOLVE identifies potential
conflicts: pairs of accesses whose byte ranges overlap where at least one is a
write.  Each potential conflict is located on two axes, the memory hierarchy,
global or shared, and the thread hierarchy, within a warp, within a CTA, or
across CTAs.  Asynchronous copies and TMA transfers are traced as well, with
each descriptor decoded into the byte ranges it moves.  When counting showed
more accesses than the buffer can hold, RESOLVE samples whole CTAs at random
rather than individual accesses, so that potential conflicts inside a CTA are
still visible.  How many CTAs to sample is a configuration knob.  A few are
usually enough, because most CTAs compute the same thing on different tiles of
data, though at least two are needed to see a conflict between CTAs.  This is
what tells RESOLVE where to place delays and at which
scope.

\begin{figure}[t]
	\centering
	\includegraphics[width=\columnwidth]{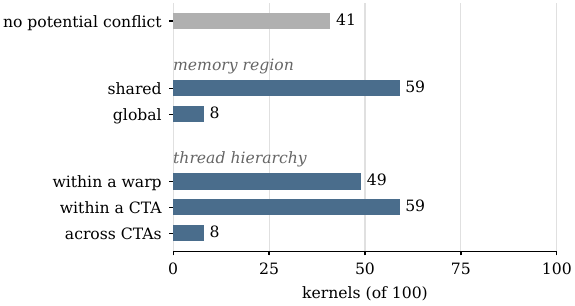}
	\caption{Where potential conflicts occur across RKB, by memory region and by
		thread hierarchy level.  Counts are out of 100 and the categories
		overlap, since a kernel can have conflicts at several levels.}
	\Description{Horizontal bar chart over 100 kernels.  41 have no potential
		conflict.  By memory region: shared 59, global 8.  By thread hierarchy
		level: within a warp 49, within a CTA 59, across CTAs 8.}
	\label{fig:conflicts}
\end{figure}

Figure~\ref{fig:conflicts} is a histogram of where potential conflicts occur
across RKB.  In 41 of the 100 kernels the trace found no potential conflict anywhere.  Those
kernels need no perturbation testing at all and are dropped from the campaign
before it starts.  The remaining 59 all have one in shared memory between
threads of the same block, making that the scope most testing has to cover.
Only 8 also have one in global memory across blocks.  That small number is not
a reason to skip the scope: all four of the issues we found in the mega-kernels
of \S\ref{sec:eval-depth} are at exactly
this scope.  It is a reason to trace every kernel rather than assume a profile,
so that testing targets the scopes that kernel actually has.

No potential conflict is reported as a bug.  RESOLVE computes no
happens-before relation, so a properly synchronized pair looks the same as an
unsynchronized one.  A potential conflict is only a place to put a delay.
Unlike a static race detector, RESOLVE reports nothing unless the delay changes
the output.

\paragraph{Does the schedule change the output?}
\label{sec:perturbation}
The third pass inserts the delays: NVBit places a call before a selected
memory instruction, and that call stalls the issuing thread for up to a set
number of cycles, with a configured probability so that not every execution is
delayed.

Perturbing a schedule to expose a concurrency bug is an old idea, and
the search has to be bounded somehow: by systematic exploration
\cite{Musuvathi2008CHESS}, by randomized scheduling with a probabilistic
guarantee \cite{Burckhardt2010PCT}, by directing it at reported races
\cite{Sen2008RaceDirected}, or on GPUs by bounding the number of delays and
directing them at conflicting atomics \cite{Chiang2013DelayBounding}.  We bound
ours by scope, using the trace to decide where a delay could matter, which keeps
the binary as the only thing we need.

RESOLVE is configured with a memory region, global or shared, and a
scope, and delays only the instructions the trace placed there.  An inter-block
delay derives the stall from the CTA index; an intra-block delay derives it
from the warp identifier.  A conflict across CTAs is exposed by separating
CTAs, not by reordering warps inside one.  How many CTAs to perturb is a
configuration knob here too.  The cycle count and probability are
parameters we fixed rather than results.  Surprisingly, neither mattered:
\S\ref{sec:eval-breadth} shows detection did not change across the ranges we
swept.

RESOLVE compares the perturbed runs against an unperturbed baseline, byte for
byte on the same input, stops at the first difference, and reports that
difference and nothing else.  Turning it into a root cause is separate work.
In \S\ref{sec:eval-depth} agents did that from these reports, and each proposed
fix was validated by running the same test again.

\paragraph{Limitations.}
RESOLVE does not perturb every scope, and the gaps follow from working at the
binary level.  We have no way to slow an asynchronous load through what NVBit
exposes.  We attempted to perturb intra-warp races and could not: every
instrumentation site is bracketed by \texttt{BSSY} and \texttt{BSYNC}, the SASS
instructions that synchronize and reconverge warp control flow, so the
independent thread scheduler never gets the chance to interleave
differently.
Conflicts between CTAs that are never resident at the same time are also out of
reach; SuperCollider shuffles block identifiers to bring such pairs together,
which needs compiler access that we do not want to require.  A clean campaign is not a
proof of race freedom, but surviving it is strong, language-agnostic,
black-box evidence that the kernel computes a deterministic function, which is
what the rest of RESOLVE needs.

\section{Reduction}
\label{sec:reduction}

Concurrency, more than size, is what puts a fully optimized GPU kernel out of
reach of a verifier.  The functional properties of asynchronous copies, TMA,
producer-consumer pipelines and the barriers coordinating them have not been
formalized, especially the corner cases that arise when using these
synchronization constructs.  Modeling each new construct as it arrives has not
kept pace with the hardware.  However, verification tools do work on much
simpler programs, so RESOLVE reduces the kernel to one of those.

Reduction runs on a candidate that passed determinism testing
(\S\ref{sec:determinism}) and whose output differs from the reference.  When
their outputs agree bitwise on every test, RESOLVE stops there: the candidate
has been tested to compute a function, and that function matches the reference.
Reduction could produce the same program from both sides, and the proof step
would add nothing.  This is not unusual, since 15 of the 100 pairs in RKB are
bitwise identical (\S\ref{sec:eval-bitwise-categories}).

An agent is instructed to rewrite each kernel using as little concurrency as
possible:

\begin{itemize}\setlength{\itemsep}{1pt}\setlength{\parskip}{0pt}
	\item Barriers, atomics, cooperative launches and warp primitives are
	      rejected, including inside headers, macros, template expansions and
	      inline PTX.
	\item Multiple kernel launches are allowed.  Splitting a large fused kernel
	      into several simple ones is usually easier than flattening it into
	      one, since we minimize concurrency and not launches.
	\item Repeated statements should become loops and shared code should become
	      functions, since the later proof effort grows with bulk.
\end{itemize}

Reduction usually reaches a single thread, but not always.  The exact
floating-point behavior of tensor cores is difficult to model in one thread,
especially across data types, and could evolve across architectures.  We
expect the future to bring more such instructions.  Rather than reduce away
from them, we allow the least concurrency the instruction requires, which for a
tensor core is a single warp, and for the warpgroup instructions on Hopper is
four.  Modern verification tools such as Kuiper
\cite{Martinez2026Kuiper} provide semantics for these instructions over the
reals, so the reduction can keep them rather than imitate them.  The floor is
set by the instruction, which is why the rule is as little concurrency as
possible and not a single thread.

A reduction is accepted only if it reproduces the output of the original,
bitwise, on every tested input.  There is no tolerance, no epsilon and no
normalization in this comparison.  We have found modern agents, with the
testing tooling available to them, very capable of producing reductions that
pass.  The result is now a simplified program that a verifier can reasonably
handle.

\section{Proof}
\label{sec:proof}
\label{sec:kplay}

After reduction, we are left with two simple programs, each bitwise equal to the
original kernel they replace.  Yet, their outputs still differ from each other.
Such a difference may be benign and not indicative of a bug: algebraically sound
changes like reassociating a reduction, distributing a product over a sum, or
using fused-multiply-add instruction instead of doing the operations
individually can change rounding behavior when using floating point numbers, and
cause low-order bits to differ. In practice, many such rewrites are routinely
done by programmers when they believe they can increase performance, as long as
some final numerical error check stays within bounds. This is unsatisfactory
since there is no principled way to define acceptable error, or which of the two
programs is more correct than the other (without using a higher-precision
reference).  We rely on testing too, but only for bitwise equality.  The problem
is the tolerance, not the testing: someone has to choose it, and that choice
decides which bugs stay visible.

In our approach, we aim to prove that the programs satisfy a common
\emph{algebraic} specification, where numerical error is ignored, and one can
only reason symbolically about the result.  That is, we do not intend to prove
that the outputs on each side are ``close enough'' (in fact we do not even want
to define what such a thing would mean) but instead prove, formally, that they
both morally compute the same algebraic expression and, that if the precision
of floats was infinite, they would match exactly and compute the exact same real
number.

This approach was introduced by Kuiper~\cite{Martinez2026Kuiper}, where the
authors introduce an abstract ``approximation'' relation ($\approx$) between
floating point numbers of any precision and real numbers. The relation can only
be introduced for literals, and is preserved by the basic arithmetic operations
(i.e. $x \approx r \wedge y \approx s$ implies $x+y \approx r+s$, where the
first $+$ is floating-point addition and the second one is real addition).
Mathematical library functions, like square root and exponentiation, get similar
assumptions. This relation decouples reasoning about the floating point numbers
and the reals they are meant to compute, such that while there is no way to
prove that $f+(g+h)$ equals $(f+g)+h$ (this would be unsound, as they can
differ), it can instead be proven that they both
approximate $x+y+z$, with $x$, $y$, $z$ reals such
that $f \approx x \wedge g \approx y \wedge h \approx z$, since,
in the reals, $(x+y)+z$ is equal to $x+(y+z)$.
This decoupling allows to give families of algorithms (e.g. GEMMs)
that make different low-level choices (due to tiling, intrinsics,
use of tensor cores, etc) a common simple top-level specification
(e.g. the output approximates the real matrix product).

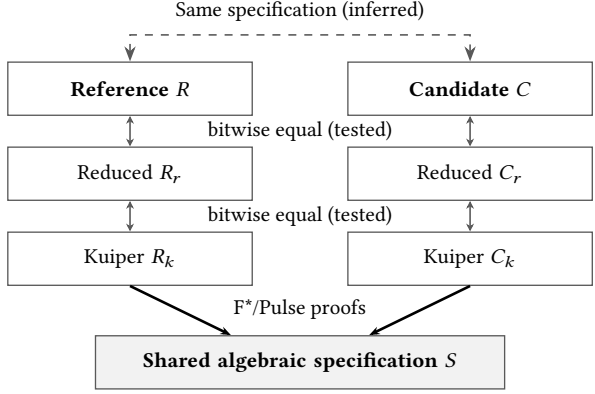
\begin{figure}[t]
	\centering
	\begin{tikzpicture}[
			font=\footnotesize,
			program/.style={draw=black!65, line width=0.5pt, align=center,
					text width=86pt, minimum height=20pt,
					inner xsep=3pt, inner ysep=2pt},
			bitwise/.style={<->, >={Stealth[length=3pt,width=3pt]},
					draw=black!65, line width=0.6pt},
			proof/.style={-{Stealth[length=4pt]}, draw=black, line width=1pt},
			inferred/.style={<->, >=Stealth, dashed, draw=black!65,
					line width=0.6pt},
			annotation/.style={font=\footnotesize, align=center},
		]
		\node[program] (base) at (-64pt,0)
			{\textbf{Reference} $R$};
		\node[program] (opt) at (64pt,0)
			{\textbf{Candidate} $C$};
		\node[program] (base-r) at (-64pt,-32pt) {Reduced $R_r$};
		\node[program] (opt-r) at (64pt,-32pt) {Reduced $C_r$};
		\node[program] (base-k) at (-64pt,-64pt) {Kuiper $R_k$};
		\node[program] (opt-k) at (64pt,-64pt) {Kuiper $C_k$};

		\draw[inferred] (base.north) -- ++(0,10pt) -| (opt.north);
		\node[annotation, anchor=south] at (0,22pt)
			{Same specification (inferred)};

		\draw[bitwise] (base.south) -- (base-r.north);
		\draw[bitwise] (opt.south) -- (opt-r.north);
		\node[annotation] at (0,-16pt) {bitwise equal (tested)};
		\draw[bitwise] (base-r.south) -- (base-k.north);
		\draw[bitwise] (opt-r.south) -- (opt-k.north);
		\node[annotation] at (0,-48pt) {bitwise equal (tested)};

		\node[program, text width=148pt, fill=black!5] (spec) at (0,-103pt)
			{\textbf{Shared algebraic specification $S$}};
		\draw[proof] (base-k.south) -- ($(spec.north)+(-26pt,0)$);
		\draw[proof] (opt-k.south) -- ($(spec.north)+(26pt,0)$);
		\node[annotation] at (0,-83pt) {F*/Pulse proofs};
	\end{tikzpicture}
	\caption{Both Kuiper programs provably satisfy $S$.  Vertical
		links test bitwise equality, using extracted CUDA for Kuiper.  The
		dashed conclusion for the originals relies on those tests.}
	\Description{Two columns, reference and candidate, each contain an original
		kernel, a reduced kernel, and a Kuiper program, from top to bottom.
		Adjacent versions in each column are linked by tested bitwise equality.
		Solid proof arrows connect both Kuiper programs to one shared algebraic
		specification.  A dashed connection between the originals denotes the
		inferred shared specification, conditional on the tested links.}
	\label{fig:proof-chain}
\end{figure}

This last feature, proving two bitwise-inequivalent programs satisfy a same
approximate specification, is the one we rely on for the last stage of RESOLVE.
For each reduced kernel, we find Kuiper implementations that are bit-equivalent
to each. This connection is also established via testing, after the Kuiper
program is extracted and built. Then, coding agents prove that the two Kuiper
programs a share a common functional approximate specification, ensuring they
perform the same algebraic computation.  Figure~\ref{fig:proof-chain} shows how
these links connect the original kernels to that specification.

To be effective at this process, it is useful to have a corpus of verified
kernels to refer to. \citet{Martinez2026KernelCorrectness} have already obtained
verified solutions to all of KernelBench level 1 challenges, and have audited
the specifications for quality to ensure, e.g., that they are not trivially
satisfiable. The agents we used during this procedure were pointed to this
public dataset and also to a growing database of kernel we had already verified.

The verification process can fail when the programs are in fact not equivalent.
In this case, the failing proof usually makes it clear where the problem is, and
coding agents can, in our experience, explain this in simple terms.
An interesting failure mode occurs when the programs output values that are
close, but that cannot be related via the approximation relation of Kuiper. Take
for instance a program computing $e^x$ using a Taylor series approximation of
degree 10, and one of degree 11. While for all practical purposes these two
could be ``close enough'' and a precise error bound could be proven, the
idealized real reasoning of Kuiper does not allow to prove these equivalent. In
the reals, both sides are simply different. Optimizations of this sort do show
up in GPU kernel optimization, and RESOLVE currently rejects them. It is
interesting future work to account for these by providing concrete error budgets
or similar.

There is a crucial component here: the specification $S$ must be \emph{tight},
i.e., even if using the approximation relation, the real numbers on the RHS must
be precisely defined. In other words, $S$ should state that the output is a
function of the inputs over the reals. It is possible in Kuiper, as in many
verification tools, to give a weak specification that only guarantees that the
program is safe, but no particular functional postcondition. If this is so, then
the fact that two programs share a common specification does not say much about
their relative behaviors. We ask the agents to construct tight specifications
only, and validate this with some human review. An automated method to certify
that specifications are tight would be useful here, and we leave it to
future work.

After a successful run, three facts now hold for each side:

\begin{itemize}\setlength{\itemsep}{1pt}\setlength{\parskip}{0pt}
	\item the kernel is bitwise equal to its reduced program;
	\item the reduced program is bitwise equal to the verified program;
	\item the verified program satisfies the common, tight specification $S$.
\end{itemize}

Both sides therefore satisfy the same specification, so the
candidate computes the function the reference defines, even though the bits
differ.

\section{Evaluation}
\label{sec:evaluation}

We evaluate RESOLVE on three sets of kernels, each fused more aggressively than
the last, ending with production mega-kernels where it exposes real bugs and
validates the agent-written fixes.

\begin{enumerate}
\item[(\S\ref{sec:eval-breadth})] On the KernelBench corpus \cite{Ouyang2025KernelBench}
we demonstrate how RESOLVE operates on a breadth of simple
kernels of the kind agents are commonly asked to
optimize.  We perform a mutation study to explore how hard concurrency errors are
to catch, and we reduce and prove a subset of the corpus.

\item[(\S\ref{sec:eval-mlp})] We examine three implementations of a
fused MLP kernel, critical to modern LLM implementations, spanning three
widely used performance-oriented languages:
CUTLASS, Gluon and Triton. For all three kernels we test, reduce and prove the
optimized variant against a PyTorch reference.

\item[(\S\ref{sec:eval-depth})] We turn to
mega-kernels, which fuse wider spans of a model into a single launch.  The
determinism tester exposes bugs in several of them, along with
other behavior that diverges significantly from their reference
implementations, and for two we show an agent repairing the kernel with the fix
tested by RESOLVE.
\end{enumerate}

The RESOLVE stages vary in how much time and tokens they consume.
The determinism tester is cheap: it samples rather than tracing everything, prunes the scopes
where a kernel can have no conflict, and works on the compiled binary.  That
leaves it fast enough to run in a tight loop, or inside a full application
wherever a deterministic output can be logged and checked
(\S\ref{sec:eval-depth}).  Reduction, specification and proof are agent tasks
and are much slower and more costly. In our context, this is acceptable,
given that the primary goal of RESOLVE is to validate kernels
as part of a larger automatic optimization system which can run without human interaction,
not an interactive tool for hunting bugs.

\subsection{Setup}
\label{sec:eval-setup}

All experiments run on a server with four NVIDIA H100 PCIe GPUs,
\texttt{sm\_90} with CUDA 13.0, hosted on an AMD EPYC 9554 running Linux 6.8.
Independent jobs take one GPU each, and no two experiments share a GPU.  Agents
appear throughout.  They perform the reductions, write the specifications,
write the Kuiper programs and check that each one reproduces the execution it
stands for, along with tasks particular to each experiment, such as mutant
generation in \S\ref{sec:eval-breadth} and kernel repair in
\S\ref{sec:eval-depth}.  Every agent runs through the Copilot CLI with
reasoning effort set to max, on GPT-5.6.

Our corpus is not
rich enough to calibrate the determinism tester empirically, so we use settings
that pilot experiments found practical: four unperturbed runs, delays of up to
5,000 cycles inserted with probability 0.8, a 200-million-record memory
tracing cap, and eight randomly chosen CTAs for both tracing and perturbation.
Section~\ref{sec:eval-breadth} shows that cap is large enough to
identify the memory regions and scopes worth testing.  Baselines are written in
PyTorch and executed in eager mode.  We assume they are deterministic, which
informal runs of our determinism tester corroborate.  All system and GPU calls
are checked for errors.

\subsection{KernelBench Level~1}
\label{sec:eval-breadth}
\label{sec:eval-bitwise-categories}

To establish how well RESOLVE works across the kinds of kernels agents are
frequently asked to optimize, we begin with a study over KernelBench Level~1 \cite{Ouyang2025KernelBench}.
We ask two questions: can the RESOLVE pipeline validate the kernels in this
suite, and when a kernel contains a bug, does RESOLVE catch it?

\paragraph{Validating kernels.}\label{sec:validating-kernels}
Recall from \S\ref{sec:setup} that RKB pairs each KernelBench Level~1 kernel
with an optimized candidate, and that each kernel comes with a set of input
distributions. Our first test over them is cheap: we run the kernel executor
and compare output hashes. 15 of the 100 candidates can be classified after
running just the determinism tester. For the remaining 85 we draw two samples
from each input distribution, and carry \nfull{} of them through the full workflow,
which keeps the study tractable.  

\begin{table}[t]
	\centering
	\caption{Outcomes over the RKB candidates (\S\ref{sec:validating-kernels}).}
	\label{tab:rkb-outcomes}
	\small
	\setlength{\tabcolsep}{4pt}
	\begin{tabularx}{\columnwidth}{@{}Xr@{}}
		\toprule
		\textbf{Outcome} & \textbf{Kernels} \\
		\midrule
				\multicolumn{2}{@{}p{\columnwidth}@{}}{\textit{Classified after first phase}} \\

		\quad Pass: Deterministic and bitwise identical & \green{\bf 15} \\
		\quad Fail: Empty kernel, no memory accesses & \green{\bf 2} \\
		\addlinespace[3pt]
		\multicolumn{2}{@{}p{\columnwidth}@{}}{\textit{Differ bitwise, require full RESOLVE workflow}} \\
		\quad Pass: Reduced and formally verified & \green{\bf 11} \\
		\quad Inconclusive: Missing Kuiper feature & \red{\bf 2} \\
		\quad Inconclusive: Numerical approximation & \red{\bf 1} \\
		\midrule
		\textbf{Total} & \textbf{31} \\
		\bottomrule
	\end{tabularx}
\end{table}

\begin{table}[t]
	\centering
	\caption{Cost per kernel over the \nfull{} RKB kernels carried through the
		full workflow.  Times are wall clock.  Medians do not sum, so the total row
		is the median over per-kernel totals.  Determinism testing (D.\ testing) is
		the only stage that uses no agent.}
	\label{tab:rkb-cost}
	\small
	\setlength{\tabcolsep}{3.5pt}
	\begin{tabular}{@{}lrrrrr@{}}
		\toprule
		& \multicolumn{3}{c}{\textbf{Time (min)}} &
		\multicolumn{2}{c}{\textbf{Median tokens}} \\
		\cmidrule(lr){2-4}\cmidrule(l){5-6}
		\textbf{Stage} & \textbf{Med.} & \textbf{Min} & \textbf{Max} &
		\textbf{In} & \textbf{Out} \\
		\midrule
		D. testing     & 1.47   & 1.05  & 19.20  & ---   & ---   \\
		Reduction      & 29.31  & 12.56 & 74.40  & 39.1K & 8.1K  \\
		Spec.\ \& proof & 112.18 & 44.55 & 250.20 & 5.54M & 65.6K \\
		\midrule
		\textbf{Total} & \textbf{142.96} & \textbf{58.16} & \textbf{343.80} &
		\textbf{5.72M} & \textbf{73.4K} \\
		\bottomrule
	\end{tabular}
\end{table}

The \textbf{top of Table~\ref{tab:rkb-outcomes}} summarizes the outcomes observed
for the 17 kernels that could be classified by just the determinism tester.
15 were found to be deterministic and bitwise equivalent. Two, however, were
detected to perform no memory accesses at all and were thus flagged by RESOLVE
as incorrect. Specifically, for problems 31 and 32, an elementwise ELU and a
hardtanh, the agent had erroneously produced an empty kernel that does no computation.  
We cannot say why, whether
it failed silently or judged its task complete too early.  Both nonetheless
pass the tolerance tests KernelBench ships with, supporting our earlier claim that
tolerance-based validation is susceptible to false negatives. 
Prior work has argued that an incorrect kernel can pass such 
thresholds \cite{Martinez2026KernelCorrectness}; these cases show an agent
producing them and a real-world tolerance suite
accepting them.

The \textbf{bottom of Table~\ref{tab:rkb-outcomes}} summarizes
the outcomes for the kernels that required running the full RESOLVE workflow.
Eleven of the \nfull{} came through.  For each, the agents reduced both
programs, wrote a specification, and produced Kuiper programs verified against
it and bitwise equivalent to the kernels they stand for. Just to be sure, we 
inspected every specification by hand and found each one to be tight.

Two of the three that did not complete the verification share a cause. Each reaches for inline
PTX to get a fast approximation or an explicit rounding mode that Kuiper has no
wrapper for: one substitutes inline assembly for \texttt{exp} and
\texttt{log1p} and flushes subnormals to zero, and another does the same for
SELU.  Wrapping thread-local operations of this kind is routine, of the sort
that added \texttt{\_\_expf} above, and nothing like the effort of formalizing
a concurrency construct.

The third is more interesting: A sigmoid candidate replaces the exponential with a
Taylor approximation, which is not algebraically equivalent to the reference. Thus
our tool "correctly" failed to prove equivalence. However, this does not mean the kernel 
is buggy. Rather, correctness of an approximating kernel simply \emph{cannot}
be determined without reasoning about numerical tolerance, of the kind developed for floating-point
expressions on CPUs \cite{Panchekha2015Herbie}, which our approach does not do
and which we leave to future work.

\textbf{Table~\ref{tab:rkb-cost}} reports what each stage cost.  Specification and proof
dominate both: they take four times as long as reduction and consume more than
a hundred times the tokens.  Almost all of that is input.  The agent reads
about eighty-five tokens for every one it writes, because each attempt at a
proof re-reads the context and returns a small edit.  Supporting these kernels
required adding \texttt{\_\_expf}
to Kuiper, which already had \texttt{expf} and takes the same specification, an
approximation of the exponential over the reals.  We expect further
thread-local semantics to be needed as RESOLVE is applied more widely.

\paragraph{Catching synchronization bugs.}\label{sec:catching-sync-bugs}
Next we investigate how RESOLVE's determinism tester fares on kernels containing
synchronization errors. In particular, we were interested to see how many and what type of
synchronization errors manifest as nondeterminism and subsequently get flagged. 
We first build a suite called \emph{RKB-Mutant} by instructing an agent to inject a single synchronization fault
into each RKB candidate, using four mutators in two groups.  

\begin{itemize}\setlength{\itemsep}{1pt}\setlength{\parskip}{0pt}
	\item \textbf{Synchronization removed, moved, or weakened.}  A barrier or
	      asynchronous wait is deleted, swapped with the memory access that
	      follows it, or narrowed in scope from cluster to block or block to
	      warp.
	\item \textbf{Write destinations aliased.}  Distinct per-thread
	      destinations are mapped onto one, so threads that never conflicted
	      now write the same location.
\end{itemize}

Applying every mutator to every kernel gives 400 attempts, which we filter in
two stages.  An agent critic discards the attempts where the mutator had no
site to apply to, such as a kernel with no barrier to remove or no shared
destination to alias, which accounts for 202.  Of the 198 that remain, a
further 55 fail to run and are excluded.  This leaves 143 mutant kernels.

The determinism tester flagged 92 of the 143 synchronization mutants.  Testing
one took 405 seconds on average, ranging from 88 to 4,587.  A catch costs less
than a clean run, because the tester stops at the first mismatch: the 92 it
caught averaged 256 seconds and none took longer than 1,091.

We reviewed all 51 misses with the help of an agent, and they fall into three
groups. Three are mutants injected into code that never runs, so there is no
fault to catch.  Twenty-eight introduce a race that never changes the value the
kernel finally writes.  The mutated kernel is still deterministic, which is the
property RESOLVE checks, so it has nothing to report, and these 28 are exactly
the case \S\ref{sec:determinism} makes when it separates a data race from a
nondeterministic result.  

The remaining 20 place their conflict inside a single
warp, where no delay we can insert will separate two threads and we know of no
perturbation an NVBit-based tool could apply to expose them.  Every mutant that
made a kernel nondeterministic and lay within reach of binary instrumentation
was caught.  Reaching inside the warp needs compiler support or a static
analysis, and is future work. Per-mutant results are given in the artifact.

Of the 92 we catch, 90 mismatch across repeated runs with no perturbation at
all. Only two need a delay to expose them.

Two cases are too few to calibrate a delay policy on, so we do not attempt it
here.  That does not mean delays are unnecessary: two of the mega-kernel
cases in \S\ref{sec:eval-depth} are exposed only under perturbation, and
prior work reports the same \cite{Stephenson2026SuperCollider}.  It is the
KernelBench mutants that do not need them, because an injected synchronization
fault usually makes the kernel visibly nondeterministic.  We swept the delay
length from 1,000 to 20,000 cycles, doubling at each step, and the probability
from 0.25 to 1 in steps of 0.25; both mutants that need a delay were detected
at every setting.

\begin{figure}[t]
	\centering
	\begin{minipage}[t]{0.48\columnwidth}
		\centering
		\includegraphics[width=\linewidth]{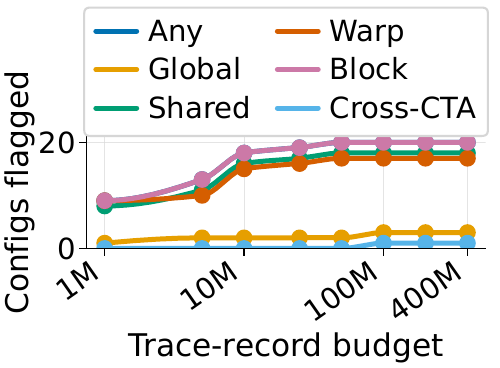}\\
		{\footnotesize\textbf{(a) Conflicts found}}
	\end{minipage}
	\hfill
	\begin{minipage}[t]{0.48\columnwidth}
		\centering
		\includegraphics[width=\linewidth]{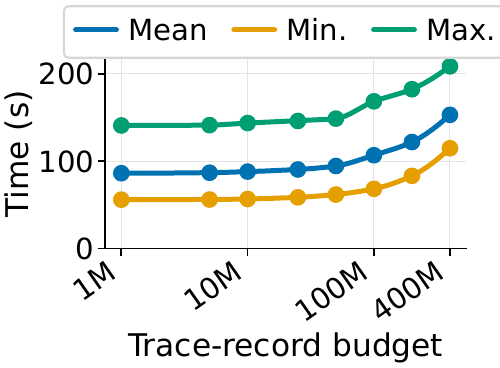}\\
		{\footnotesize\textbf{(b) Time to trace}}
	\end{minipage}
	\caption{Choosing the trace record budget.  (a) Potential conflicts found,
		by memory region and by thread hierarchy level.  The inventory is
		complete at 100 million records and unchanged at 200 and 400 million.
		(b) Mean, minimum and maximum time to trace and analyze one kernel,
		which keeps rising past that point.  By default we trace 200 million
		records.}
	\Description{Two line charts against trace record budget from one million
		to four hundred million on a logarithmic axis.  The left chart plots six
		conflict indicator counts, which rise and then flatten from one hundred
		million onward.  The right chart plots mean, minimum and maximum time to
		configure a kernel, which continues to rise across the same range.}
	\label{fig:trace-budget}
\end{figure}

Even where perturbation is not required, the tracing pass still has a budget,
and we can measure both what it costs and how many potential conflicts it
finds.  \textbf{Figure~\ref{fig:trace-budget}} sweeps that budget from one million
records to four hundred million.  Below 100 million the trace saturates on the
larger kernels and loses conflicts entirely: at 50 million the inter-block
indicator disappears.  At 100 million every case traces to completion, and the
conflict inventory is unchanged at 200 and 400 million while the mean time to
configure a kernel rises from 107 to 153 seconds.  We trace 200 million
records, one step past the point where the inventory stops changing.  Two CTAs
were enough to find every conflict indicator on this suite, and two is the
fewest that can expose a conflict between CTAs at all.  That nothing further
appears beyond two is evidence for the assumption in \S\ref{sec:tracing} that
CTAs are largely interchangeable.

\subsection{A Fused MLP Kernel}
\label{sec:eval-mlp}

The kernel fuses the up projection, the gate projection and the SwiGLU
activation of an MLP block, the arrangement used by current models such as
Qwen.  Written the usual way this is three kernels, and the two projections
must write their results to memory for the activation to read them back.
Fusing the three removes four passes over those intermediates and makes the
block run up to $2.5\times$ faster than the PyTorch implementation of the same
block.  Neither PyTorch's compiler nor TensorRT performs the fusion, because it
has to cross two projections that use different weight matrices.  At the time of
writing, no compiler will automatically produce this kernel.

Our reference is PyTorch in eager mode, where the three operations run as three
kernels.  Against it we place three aggressively optimized variants, one each
in Triton \cite{Tillet2019Triton}, Gluon \cite{Triton2026Gluon} and CUTLASS
\cite{NVIDIA2023CUTLASS}, each fusing the work into a single kernel and each
written with the help of an agent.  They are the fastest implementations
of this operation we are aware of.  We validate them at a representative size,
a hidden dimension of 8,192 and an FFN intermediate size of 29,568 against
4,096 rows of activations.  None of the three matches the reference bitwise,
because the fusion reorders the reductions, and all three are intricate enough
that correctness cannot be taken for granted.  Gluon and Triton do match each
other bitwise, so we carried only Gluon through specification and proof.

As far as we are aware, these kernels are out of reach for existing rigorous
validation, on both halves of the problem.  For concurrency, Compute Sanitizer
\cite{NVIDIA2026ComputeSanitizer} runs on all three and reports nothing, which
is the answer we expect and also all it can offer: a clean race report says
neither that the kernel is deterministic nor that it computes the reference
function.  SuperCollider \cite{Stephenson2026SuperCollider} reaches individual
accesses through a CUDA compiler pass, which two of our three variants cannot go
through because they are not written in CUDA.  For functional correctness, the
kernels are built out of asynchronous memory copies and tensor core
instructions, which verification tools such as ProofWright
\cite{Chatterjee2026ProofWright} do not model.

RESOLVE establishes the equivalence of all three variants against the PyTorch
reference.  Table~\ref{tab:mlp-cost} reports what each one took.  Reduction and
proof both run faster than the RKB medians in Table~\ref{tab:rkb-cost}, even
though these kernels are far more intricate, because matrix multiplication was
studied heavily in the work that produced Kuiper~\cite{Martinez2026Kuiper} and
the corpus an agent can draw on is correspondingly rich for kernels of this
shape.  Determinism testing runs the other way and takes several times longer,
because these kernels run at the sizes a deployed model uses rather than
benchmark sizes, leaving far more memory traffic to trace.

New languages and compilers for writing this kind of code keep
appearing~\cite{NVIDIA2026CuTile,Wang2025TileLang,Ye2026CAKE,Soi2026Twill}, and
which one produces the fastest kernel for a given operation keeps changing as
they develop.  An agent optimizing a kernel
should be able to move it from any one of these languages to another and still
know it computes the same function, which is what RESOLVE gives it.

\begin{table}[t]
	\centering
	\caption{Time to validate each variant of the fused MLP kernel, in minutes.
		Gluon and Triton are bitwise identical, so only Gluon was specified and
		proved.  Starred entries are estimates at about an hour each, since a
		logging fault on those runs lost the exact time.}
	\label{tab:mlp-cost}
	\small
	\setlength{\tabcolsep}{6pt}
	\begin{tabular}{@{}lrrr@{}}
		\toprule
		\textbf{Stage} & \textbf{CUTLASS} & \textbf{Gluon} & \textbf{Triton} \\
		\midrule
		D.\ testing     & 9.73  & 8.18  & 10.40 \\
		Reduction       & 17.17 & 15.56 & 10.92 \\
		Spec.\ \& proof & ${\sim}60^{*}$ & ${\sim}60^{*}$ & --- \\
		\bottomrule
	\end{tabular}
\end{table}

\subsection{Mega-Kernels}
\label{sec:eval-depth}

A mega-kernel takes fusion further again.  Where \S\ref{sec:eval-mlp} fuses
three operations into one kernel, these fuse much wider spans of a model, and
they are the most aggressively fused GPU code we know of.  We ran RESOLVE over
three of them, each in its single-GPU configuration.

\begin{itemize}\setlength{\itemsep}{1pt}\setlength{\parskip}{0pt}
	\item \textbf{LUCE}~\cite{Ciffa2026Luce} fuses all 24 decoder layers of a
	      hybrid DeltaNet and attention model into a single launch.
	\item \textbf{FlashMoE}~\cite{Aimuyo2025FlashMoE} fuses the routing and
	      expert dispatch of a mixture-of-experts layer.
	\item \textbf{FastMoE}~\cite{He2021FastMoE} provides a native
	      capacity-pruning kernel for mixture-of-experts routing.
\end{itemize}

We also attempted Mirage~\cite{Cheng2025MPK}, which compiles a model into a
persistent mega-kernel.  It hung when run under our determinism tester and we
could not diagnose why, so we report nothing for it.  Per-system configurations
are given in the artifact.

\paragraph{Testing mega-kernels.}
These kernels differ from everything above in one practical respect: they do
not come with kernel executors.  Each is embedded in a larger framework and is
exercised only through that framework's own tests.  Our first pass therefore
runs the determinism tester over those existing test suites, with an agent
identifying outputs that should be deterministic and logging them at each stage
of the pipeline.

That the determinism tester can be pointed at code in this shape is a
consequence of working at the binary.  It is at its most convenient given a
kernel executor, but NVBit will instrument nearly any NVIDIA binary, so the
requirement reduces to identifying an output that should not change between
runs and logging it.  Nothing about the framework around the kernel has to be
understood.

What each run exercises, and what was logged from it:

\begin{itemize}\setlength{\itemsep}{1pt}\setlength{\parskip}{0pt}
	\item \textbf{LUCE}: a full decode step of Qwen3.5-0.8B.
	      \emph{Logged:} the generated token and the final hidden state.
	\item \textbf{FlashMoE}: the router alone, with the experts overloaded so
	      the capacity limit binds.  \emph{Logged:} expert demand, and the
	      surviving tokens and their probabilities.
	\item \textbf{FastMoE}: the capacity-pruning kernel, on assignments that
	      exceed capacity.  \emph{Logged:} the pruned experts and the survival
	      mask.
\end{itemize}

\paragraph{Nondeterministic outputs.}
All three mega-kernels behaved nondeterministically.  An agent could then work on
each one with the determinism tester in hand, guessing at a cause, re-running the
tester to check the guess, and narrowing the search by delta
debugging~\cite{Zeller2002DeltaDebugging}.  Four root causes came out: two in
LUCE, one in FlashMoE and one in FastMoE.

LUCE's two are plainly bugs.  In the grid barrier, block 0 resets the arrival
counters while other blocks may already be arriving, and
\texttt{\_\_syncthreads()} cannot order one block against another.  The blog post
that introduces the kernel mentions hitting grid synchronization problems during
development~\cite{Ciffa2026Luce}, so a barrier problem is not a complete
surprise.  The second one looks unknown: several attention blocks write the same
global scratch buffer, because the address they compute for it carries no block
or head offset, and nothing we found mentions it.
The barrier was also the only one that showed up reliably without perturbation. The other three needed inserted
delays before they appeared consistently.

The two routing causes are harder to classify.  Both come from capacity.  When an
expert fills, routed tokens are dropped, and in both systems the tokens that get
dropped depend on the order blocks arrive: FlashMoE reserves slots with atomics,
and FastMoE prunes with \texttt{atomicSub}.  Every reference implementation we
know of routes deterministically and drops the same set of tokens once capacity
binds.  These two do not, and neither paper says so.  Whether that should be
treated as a bug is under discussion.

\paragraph{Fixing the bugs.}
We asked the agents to repair each root cause.  A kernel counted as fixed once it
ran under the determinism tester without any nondeterminism showing up, which is
a weaker bar than the rest of this paper uses.
Table~\ref{tab:megakernel-fixes} gives what each repair changed and what it
cost in performance.  The two LUCE repairs came out marginally faster than the code they
replaced, which we read as noise rather than a win.  FlashMoE's costs under a
percent, so there is no performance argument for leaving its router
nondeterministic.  FastMoE is the one case where determinism is not close to
free, at ten percent, and that is the case where whether the behavior is a bug
is worth a discussion.

We then attempted the whole pipeline on each fixed kernel, writing a
straightforward PyTorch reference out of the operators the system implements.  FastMoE came out bitwise
identical to its reference, so nothing further was needed.  FlashMoE did not.
We reduced it in 25 minutes, but could not specify it in Kuiper, which has no
support for argmax.  The router picks experts with one, and an argmax returns an
index rather than a real and is discontinuous at ties, so the approximation
relation of \S\ref{sec:proof} does not carry across it.  Specifying the capacity
stage on its own, taking the assignments as input, avoids the argmax, and we
leave this to future work.  LUCE we
could not reduce at all: its matrix multiplies use the
Hopper warpgroup instructions, whose floor is 128 threads, and our reducer
handles nothing above a single warp.  Raising that floor means extending the
reducer, and extending Kuiper with those semantics as well.

We are working on reporting each of these upstream, along with the fix we
recommend.

\begin{table}[t]
	\centering
	\caption{What each repair changed, and what it cost.  Latency is the change
		in warmed call latency after the repair, so a negative number is faster.
		The two LUCE rows compare each original variant against the combined
		repair.}
	\label{tab:megakernel-fixes}
	\small
	\setlength{\tabcolsep}{4pt}
	\begin{tabularx}{\columnwidth}{@{}l>{\raggedright\arraybackslash}Xr@{}}
		\toprule
		\textbf{Root cause} & \textbf{Repair} & \textbf{Latency} \\
		\midrule
		LUCE (barrier) & Order the counter reset against arriving blocks &
		$-1.4\%$ \\
		LUCE (scratch) & Give the scratch address a block and head offset &
		$-1.6\%$ \\
		FlashMoE & Reserve slots independently of arrival order & $+0.8\%$ \\
		FastMoE & Prune capacity independently of arrival order & $+10.5\%$ \\
		\bottomrule
	\end{tabularx}
\end{table}

\section{Related Work}
\label{sec:related}

\paragraph{Determinism and race freedom.}
Data-race freedom buys a simpler model to reason in, since a program without
races behaves as though the machine were sequentially
consistent~\cite{Adve1990WeakOrdering}.  Determinism buys something similar and
has its own history, going back to determinacy races in
Cilk~\cite{Feng1997DeterminacyRaces} and to runtime checks that a program's
output does not depend on the schedule~\cite{Burnim2009Determinism}.  Type
systems can guarantee determinism statically, on the CPU~\cite{Bocchino2009DPJ} and on
the GPU~\cite{Kopcke2023Descend}.

Lipton reduction merges statements into coarser atomic
blocks~\cite{Lipton1975Reduction,Elmas2009QED}.  Sequentialization goes further
and emits an equivalent sequential
program~\cite{Lal2008Sequentialization,Inverso2014LazySeq}.  These are complete
guarantees, but require source code to be available.  By contrast, we run the
binary under perturbation and watch the output; \S\ref{sec:eval-breadth} lists
the mutants that slip past.  However our approach scales to binary kernel
compiled from in Gluon or CUTLASS.

\paragraph{Verified optimizers.}
Tools like CompCert and Exo provide general correctness guarantees, one by
verifying the compiler~\cite{Leroy2006CompCert} and the other by building
schedules only out of transformations already known to preserve
semantics~\cite{Ikarashi2022Exo}.

A long line of work checks that the output of the optimizer is equivalent to
its input for each particular run, instead of trusting or verifying the tool.
Translation validation checks a single compilation instead of the
compiler~\cite{Pnueli1998TranslationValidation,Necula2000TranslationValidation}.
Alive2 takes two LLVM functions and determines refinement by encoding the
problem to SMT~\cite{Lopes2021Alive2}.
This approach has also been used in superoptimizers to ensure
correctness~\cite{Schkufza2013STOKE}.  For tensor programs the same problem
appears as
graph substitutions, either verified up front~\cite{Jia2019TASO} or allowed to
be wrong and patched afterwards~\cite{Wang2021PET}.

RESOLVE by contrast takes an even more radical approach: not only do we check
each problem individually, we also pick a separate notion of equivalence for
each problem (by deriving a natural spec that both programs are verified
against).

\paragraph{GPU semantics.}
Formal models of GPU behavior tend to show up well after the hardware.  A first
memory consistency model was pinned down empirically in
2015~\cite{Alglave2015GPUConcurrency}.  NVIDIA published a formal one four years
after that~\cite{Lustig2019PTXMemoryModel}.  Testing hardware against such models
at scale still finds violations~\cite{Levine2023GPUHarbor}.  Execution inside a warp has no
official model, only a proposed operational semantics~\cite{Chen2026SIMTStep}.
Forward progress between workgroups is undefined, and the candidate models had
to be run on real hardware to see which ones
held~\cite{Sorensen2021ForwardProgress}.  Named barriers got a model built to
verify them~\cite{Sharma2015WEFT}.  Tensor core arithmetic has an SMT
formalization, though not from NVIDIA~\cite{Valpey2025TensorCoreSMT}.  Scoped
data-race freedom is a cleaner foundation~\cite{Hower2014HRF}, but it only
covers programs that respect the scopes, which an aggressively tuned kernel may
not.

None of this blocks RESOLVE's determinism tester, since it does not require a
formal model of each hardware feature. It only tests the general property that
how long a memory access takes should not change the semantics of a program.
Therefore we can verify kernels that use named barriers without even needing to
know what named barriers are, let alone have a correct and practically usable
formal semantics for them. Kuiper however still requires semantics for features
that are used in the sequentialized kernels, and that blocks two kernels using
certain thread-local operations (\S\ref{sec:eval-breadth}).

\section{Conclusion}

RESOLVE combines testing, reduction, and proof to validate optimized GPU
kernels across languages and frameworks. Our evaluation evaluated 31
KernelBench Level~1 kernels, three fused MLP implementations in CUTLASS,
Triton, and Gluon against a common PyTorch reference, and a
mixture-of-experts kernel from an existing system. The determinism tester
also caught 92 injected synchronization faults and exposed four previously
unreported issues in production mega-kernels, including two clear bugs.
These results show that the approach can both uncover errors in deployed
code and support validation of optimizations beyond the verifier's direct
semantic coverage.  As agents take on more of kernel development, they need
validation they can trust, enough detail to debug what fails, and the freedom to
write in whatever language is fastest at the time.

\section*{Acknowledgments}
Agents were used widely throughout this work.  They wrote and optimized kernels,
performed the reductions, drafted specifications and Kuiper proofs, and proposed
the repairs.  The text of this paper was organized and drafted by people, and
agents polished and reviewed it.  The
project itself was directed by people.  The scientific arguments, the validation
of every result reported here, and the integrity of the artifact are ours.  The
agents ran through GitHub Copilot CLI, which lets the underlying model be configured; most
of this work used GPT-5.6 and Claude Opus 5.

\bibliographystyle{ACM-Reference-Format}
\bibliography{reference}

\end{document}